\documentclass[aps,pra,letterpaper,twocolumn,showpacs,amsmath,amssymb,superscriptaddress]{revtex4}
\usepackage{graphicx}
\usepackage{setspace}
\usepackage{dcolumn}
\usepackage{mathrsfs}
\usepackage{amssymb}
\usepackage{mathtools}
\usepackage{bm}
\usepackage{longtable}
\usepackage{xcolor}
\usepackage{braket}
\usepackage{hyperref} % For hyperlinks in the PDF
\hypersetup{
	colorlinks=true,
	linkcolor=blue,
	citecolor=red,
	urlcolor=blue
}

\begin{document}
\preprint{}

\title{Dynamics of position disordered Ising spins with a soft-core potential}

\author{Canzhu Tan}\author{Xiaodong Lin}\author{Yabing Zhou}
\affiliation{Hefei National Laboratory for Physical Sciences at the Microscale and Shanghai Branch, University of Science and Technology of China, Shanghai 201315, China}
\affiliation{CAS Center for Excellence and Synergetic Innovation Center in Quantum Information and Quantum Physics, University of Science and Technology of China, Shanghai 201315, China}

\author{Y. H. Jiang}

\affiliation{Shanghai Advanced Research Institute, Chinese Academy of Sciences, Shanghai 201210, China}
\affiliation{CAS Center for Excellence and Synergetic Innovation Center in Quantum Information and Quantum Physics, University of Science and Technology of China, Shanghai 201315, China}

\author{Matthias Weidem\"uller}

\affiliation{Hefei National Laboratory for Physical Sciences at the Microscale and Shanghai Branch, University of Science and Technology of China, Shanghai 201315, China}
\affiliation{CAS Center for Excellence and Synergetic Innovation Center in Quantum Information and Quantum Physics, University of Science and Technology of China, Shanghai 201315, China}
\affiliation{Physikalisches Institut, Universit\"at Heidelberg, Im Neuenheimer Feld 226, 69120 Heidelberg, Germany}

\author{Bing Zhu}
\email{bzhu@physi.uni-heidelberg.de}
\affiliation{Physikalisches Institut, Universit\"at Heidelberg, Im Neuenheimer Feld 226, 69120 Heidelberg, Germany}
\affiliation{CAS Center for Excellence and Synergetic Innovation Center in Quantum Information and Quantum Physics, University of Science and Technology of China, Shanghai 201315, China}

\date{\today}

\begin{abstract}
	We theoretically study magnetization relaxation of Ising spins distributed randomly in a $d$-dimension homogeneous and Gaussian profile under a soft-core two-body interaction potential $\propto1/[1+(r/R_c)^\alpha]$ ($\alpha\ge d$), where $r$ is the inter-spin distance and $R_c$ is the soft-core radius. The dynamics starts with all spins polarized in the transverse direction. In the homogeneous case, an analytic expression is derived at the thermodynamic limit, which starts as $\propto\exp(-kt^2)$ with a constant $k$ and follows a stretched-exponential law at long time with an exponent $\beta=d/\alpha$. In between an oscillating behaviour is observed with a damping amplitude. For Gaussian samples, the degree of disorder in the system can be controlled by the ratio $l_\rho/R_c$ with $l_\rho$ the mean inter-spin distance and the magnetization dynamics is investigated numerically. In the limit of $l_\rho/R_c\ll1$, a coherent many-body dynamics is recovered for the total magnetization despite of the position disorder of spins. In the opposite limit of $l_\rho/R_c\gg1$, a similar dynamics as that in the homogeneous case emerges at later time after a initial fast decay of the magnetization. We obtain a stretched exponent of $\beta\approx0.18$ for the asymptotic evolution with $d=3, \alpha=6$, which is different from that in the homogeneous case ($\beta=0.5$).
\end{abstract}

\maketitle

\section{Introduction}

Disorder plays an essential role in determining both equilibrium and non-equilibrium properties of a many-body system, e.g. glassy phase and dynamics in magnetics \cite{Binder1986}, localization phenomenon of transports \cite{Muller2016}, and novel materials by disorder engineering \cite{Upadhyaya2018, Meier2018, Yu2021}. While knowing very details of a disordered system is difficult and not necessary, understanding its universal behaviour starting from a microscopic Hamiltonian is important to pin down the underlying physics. For example, many relaxations in glass materials (normal or spin-type) follow a simple stretched-exponential law ($\propto\exp[-(\gamma t)^\beta], \beta<1$) \cite{Binder1986, Phillips1996}. Klafter and Shlesinger found that a scale-invariant distribution of relaxation times was the common underlying structure for three different physical models showing a stretched-exponential decay \cite{Klafter1986}, which was generalized to closed quantum systems by Schultzen and coworkers recently \cite{Schultzen2021a}.

For a disordered spin-1/2 system, recent studies confirmed a stretched-exponential decay of magnetization in both Ising \cite{Schultzen2021a} and Heisenberg \cite{Signoles2021a, Schultzen2021} models, where the pairwise spin-spin interaction exhibits a pow-law dependence on the inter-spin distance $r$, $J(r)\propto1/r^\alpha$ with $\alpha\ge d$ in the $d$-dimension. The scale invariance is guaranteed since pairwise contribution to the relaxation dynamics is determined by $J(r)t$, which is invariant under the following rescaling of space and time: $r\rightarrow\lambda r$ and $t\rightarrow\lambda^\alpha t$. 

How would the stretched-exponential law change if the scale invariance is broken? Here we consider a specific type of pairwise interactions in an Ising Hamiltonian, namely a soft-core potential $J(r)\propto1/[1+(r/R_c)^\alpha]$ with $\alpha\ge d$ and $R_c$ the soft-core radius, reducing to the power-law behaviour at large $r$. We have studied two different situations: (i) For homogeneously distributed spins, an analytical formula is derived for the magnetization relaxation at the thermodynamic limit, which features three different regions in the time axis: The dynamics starts as $\propto\exp(-kt^2)$, followed by an oscillating decay, and eventually obeys an stretched-exponential law. (ii) For a spatially inhomogeneous sample, e.g. Gaussian distributed, a coherent many-body dynamics is observed in small-spatial-size system while disorder-induced relaxation is recovered for large spatial sizes. Our investigation concerning the soft-core potential is inspired by Rydberg dressing in cold-atom experiments (for recent reviews see \cite{Balewski2014, Browaeys2020}) and both studied situations can be readily tested there; A uniform gas can be prepared via box potentials \cite{Gaunt2013, Mukherjee2017} and a Gaussian distribution of atoms is obtained with a harmonic trap \cite{Ketterle1999}.  

The article is organized as follows: We derive and discuss the analytical result for homogeneous samples in Sec. \ref{sec:homo}. The inhomogeneous situation is numerically investigated in Sec. \ref{sec:inhomo} and Sec. \ref{sec:conclusion} concludes the paper.

\section{Disordered Ising model with a soft-core potential} \label{sec:dynamics}

\subsection{The Ising Hamiltonian and its dynamics} \label{sec:hamiltonian}

\begin{figure}[t]
	\includegraphics[width=0.5\textwidth]{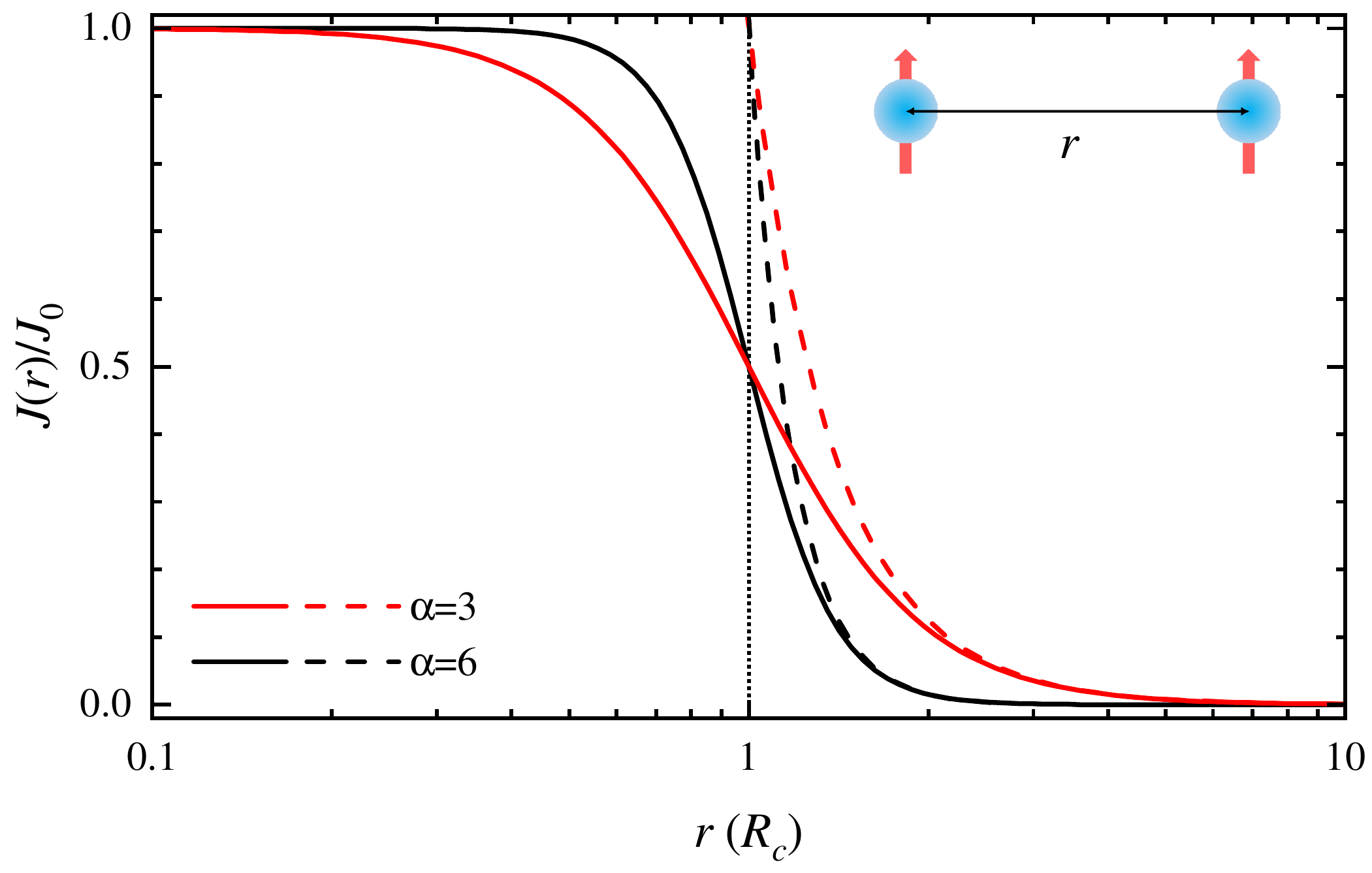}
	\centering
	\caption{Soft-core interaction potential between two spin-up particles. The soft-core potential in Eq. \eqref{eq:Vr} is plotted as a function of inter-spin distance $r$ with $\alpha=3$ and 6 (solid curves). As a comparison, the power-law interactions $J_0/(r/R_c)^\alpha$ are also shown.}
	\label{fig:dressing_interaction}
\end{figure}

A general Ising Hamiltonian for $N$ spin-1/2 particles reads
\begin{equation} \label{eq:Ising} % Example of optimization equation
	\hat{H}_{\text{Ising}} = \frac{1}{2}\sum_{i,j}^{N}J_{ij}\hat{\sigma}_i^z \hat{\sigma}_j^z \, ,
\end{equation}
where $\hat{\sigma}_{i(j)}^z$ is the Pauli $z$ operator and $J_{ij}$ is the coupling strength between spins $i$ and $j$. $J_{ij}$ takes a form of the soft-core potential
\begin{equation} \label{eq:Vr}
	J_{ij}\equiv J(r_{ij}) = \frac{J_{0}}{1+(\frac{r_{ij}}{R_{c}})^\alpha} \, ,
\end{equation}    
where the long-range part ($r_{ij}\gg R_c$) has a power-law form ($\propto 1/r_{ij}^\alpha$) and the short-range ($r_{ij}\ll R_c$) interaction is almost a constant $J_0$, as seen in Fig. \ref{fig:dressing_interaction}. Such a potential is not invariant under the spatial scaling $r\rightarrow \lambda r$ in general, while it is approximately invariant at large $r\gg R_c$. We will see later that this leads to a stretched-exponential relaxation for long-time dynamics both in the analytic solution of a homogeneous sample and the numerical results of a Gaussian one.

We focus on dynamics of the mean magnetization $\braket{\hat{S}_x(t)}=\braket{\sum_{i=0}^N \hat{\sigma}_x^i(t)}/N$ with a initial state that all spins are polarized in the $+x$ direction $\ket{\phi_0}=\ket{\rightarrow}^{\bigotimes N}$ with $\hat{\sigma}_x\ket{\rightarrow}=+1\ket{\rightarrow}$, i.e. $\braket{\hat{S}_x(0)}=1$. Emch \cite{Emch1966} and Radin \cite{Radin1970} have obtained an analytical expression for $\braket{\hat{S}_x(t)}$ with the initial state $\ket{\phi_0}$, which reads as
\begin{equation} \label{eq:EmchRadin}
	\braket{\hat{S}_x(t)}=\sum\limits_{i=1}^N 1/N\prod_{j\neq i}\cos(J_{ij}t) \, .
\end{equation}
All following analytical and numerical results are based on the above equation.

\subsection{Homogeneous samples: the thermodynamic limit} \label{sec:homo}

\begin{figure}[t]
	\includegraphics[width=0.5\textwidth]{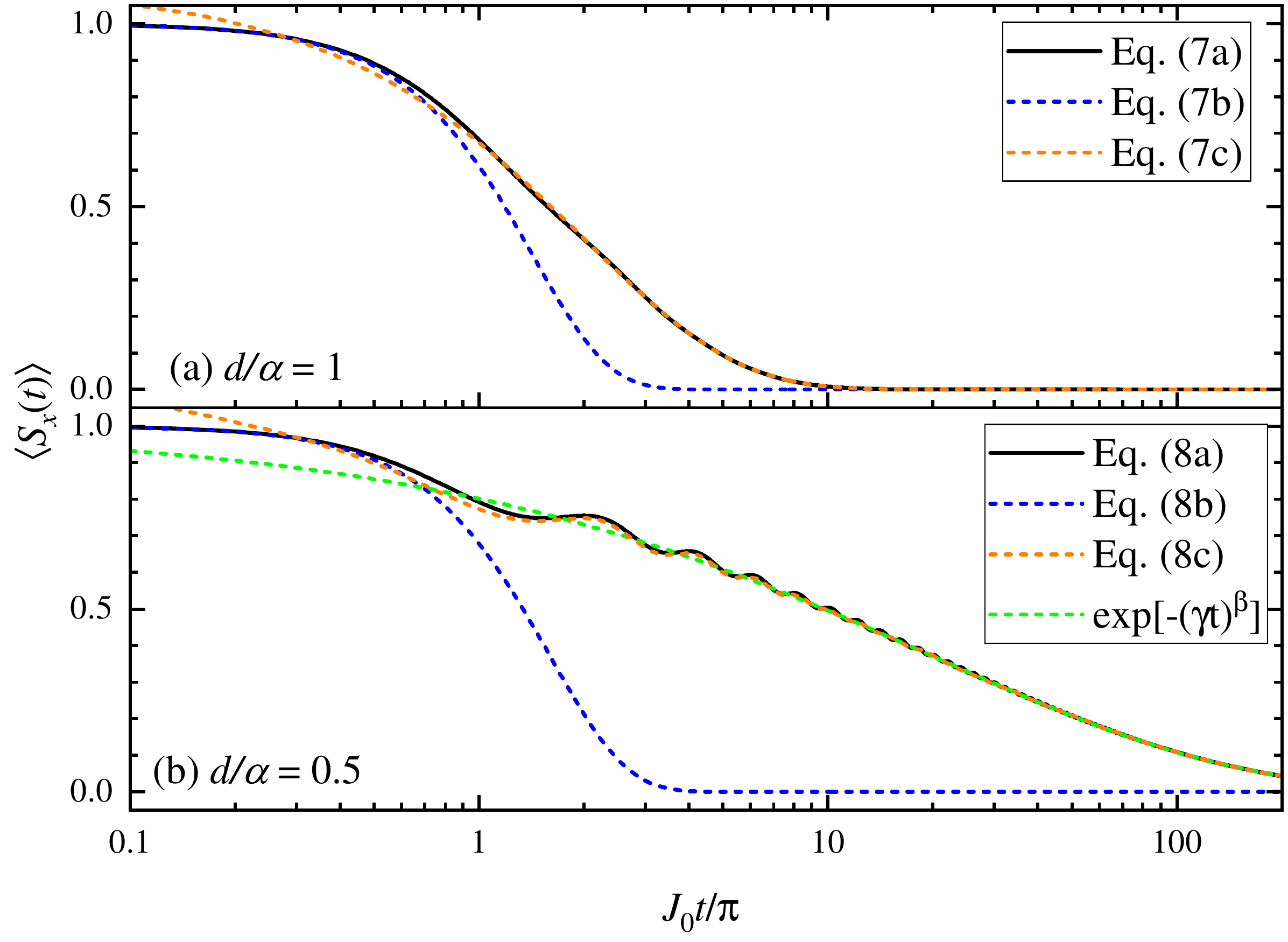}
	\centering
	\caption{Plots of analytic results for magnetization relaxation in uniformly distributed Ising spins under a pairwise soft-core interaction. (a) The case of $\beta_0=d/\alpha=1$. The black solid, blue dashed, and orange dashed curves represent the complete [Eq. \eqref{eq:int1a}], short-time asymptotic [Eq. \eqref{eq:int1b}], and long-time asymptotic [Eq. \eqref{eq:int1c}] expressions, respectively. (b) Similar as in (a) for the case of $\beta_0=d/\alpha=0.5$. We also show the stretched-exponential function as a green curve as comparison.}
	\label{fig:homo}
\end{figure}

We consider a system of $N$ spins uniformly distributed in a spherical volume $V$ in the $d$ dimension. Following the same derivation procedure in Ref. \cite{Schultzen2021a}, by replacing the ensemble average with an average over all possible configurations of placing $N-1$ spins around a reference one at $\mathbf{r}_1=\mathbf{0}$, Eq. \eqref{eq:EmchRadin} can be transformed to 
\begin{equation} \label{eq:homo_exp}
	\begin{aligned}
		\braket{\hat{S}_x(t)} & =\int_V d\mathbf{r}_2\cdots d\mathbf{r}_N P(\mathbf{r}_2, \cdots, \mathbf{r}_N) \prod\limits_{j=2}^N\cos(J_{1j}t) \\
		& = \{\frac{1}{V}\int_V d\mathbf{r}\cos[J(r)t]\}^{N-1} \\
		& = \{\frac{d}{r_0^d}\int_0^{r_0} r^{d-1}dr\cos[\frac{J_0t}{1+(r/R_c)^\alpha}]\}^{N-1}
	\end{aligned}
	 \, .
\end{equation} 
Here $P(\mathbf{r}_2, \cdots, \mathbf{r}_N)=1/V^{N-1}$ is the probability of placing the $N-1$ spins at positions $\mathbf{r}_2, \cdots, \mathbf{r}_N$, respectively, and $J(r)$ takes the form in Eq. \eqref{eq:Vr}.

For the power-law interaction ($\propto1/r^\alpha$) a short-distance cutoff has to be introduced to avoid the divergence of interaction strength for further simplifying Eq. \eqref{eq:homo_exp} \cite{Schultzen2021a}, which is not necessary for the soft-core potential considered here. By introducing a new variable $y=J_0t/[1+(r/Rc)^\alpha]$ and integrating by parts, Eq. \eqref{eq:homo_exp} can be written as
\begin{equation} \label{eq:homo_exp1}
	\begin{aligned}
		\braket{\hat{S}_x(t)} & = [1-\frac{\pi^{d/2}\rho R_c^d}{\Gamma(d/2+1)N} \int_{y_0}^{J_0t} (J_0t/y-1)^{\beta_0}\sin y dy]^{N-1}
	\end{aligned}
	\, ,
\end{equation}
where $N=\rho V =\rho \pi^{d/2} r_0^d/\Gamma(d/2+1)$, $\beta_0=d/\alpha$, and $y_0=J_0t/[1+(r_0/Rc)^\alpha]$. Here $\rho$ is the particle density and $\Gamma(x)$ is the Gamma function. In the thermodynamic limit ($N, r_0\rightarrow \infty$ and $\rho$ is a constant), the integral $I(J_0 t; \beta_0)=\int_{y_0}^{J_0t} (J_0t/y-1)^{\beta_0}\sin y dy$ is finite only if $\beta_0\le1$ and the above equation gives 
\begin{equation} \label{eq:homo_exp2}
	\begin{aligned}
		\braket{\hat{S}_x(t)} & = \exp[-FI(J_0t;\beta_0)]
	\end{aligned}
	\, ,
\end{equation}
where $F=\pi^{d/2}\rho R_c^d/\Gamma(d/2+1)$.

Let us first consider $\beta_0=1$, 
\begin{subequations}
\begin{align}
	I(J_0t;1) & = J_0t\text{Si}(J_0t)+\cos(J_0t)-1 \, , \label{eq:int1a} \\
	\intertext{where $\text{Si}(x)=\int_0^x\sin(t)/t dt$ is the sine integral function. At short times ($J_0t\ll1$),}
	I(J_0t;1) & \sim1/2(J_0t)^2 \, ,  \label{eq:int1b}\\
	\intertext{while}
	I(J_0t;1) & \sim\pi/2J_0t-1 \label{eq:int1c}
\end{align}
\end{subequations}
for long times ($J_0t\gg1$). In Fig. \hyperref[fig:homo]{2(a)}, all three formulae are plotted as a function of evolution time and Eqs. \eqref{eq:int1b} and \eqref{eq:int1c} describe excellently the asymptotic dynamics at short and long times, respectively. Specifically, a stretched-exponential decay, $\exp[-(\gamma t)^\beta]$ with $\beta=\beta_0=1, \gamma=J_0F\pi/2$, is seen for the long-time dynamics.  

Next we look at $\beta_0<1$, the integral $I(J_0 t; \beta_0)$ is
\begin{subequations}
	\begin{equation}\label{eq:int2a}
		I(J_0t;\beta_0) = J_0t B(1-\beta_0, 1+\beta_0) \Im[_1F_1(1-\beta_0, 2, iJ_0t)] \, , 
	\end{equation}
where $B(x,y)$ is the Euler beta function, $_1F_1(a, b, z)$ is the Kummer confluent hypergeometric function, and $\Im[z]$ gives the imaginary part of $z$. More details can be found in Appendix \ref{app:ad_homo}. The asymptotic behaviors of Eq. \eqref{eq:int2a} are
		\begin{equation}
			I(J_0t;\beta_0) \sim1/2(J_0t)^2(1-\beta_0)B(1-\beta_0, 1+\beta_0)  \label{eq:int2b}
		\end{equation}
for short times ($J_ot\ll1$), and
		%I(J_0t;\beta_0) & \sim (J_0t)^{\beta_0}\cos(\beta_0 %\pi/2)\Gamma(1-\beta_0)  \label{eq:int2c}
		\begin{widetext}
			\begin{equation}
				I(J_0t;\beta_0) \sim (J_0t)^{\beta_0}[\cos(\beta_0 \pi/2)\Gamma(1-\beta_0)-(J_0t)^{-2\beta_0}\cos(J_0t-\beta_0\pi/2)\Gamma(1+\beta_0)] \label{eq:int2c}
			\end{equation}
		\end{widetext}
\end{subequations}
for long times ($J_0t\gg1$).

The asymptotic form of Eq. \eqref{eq:int2b} for $\beta_0\rightarrow1$ actually coincides with Eq. \eqref{eq:int1b}. Thus, for $\beta_0\le1$ the initial dynamics of $\braket{\hat{S}_x(t)}$ follows $\exp[-kt^2]$ with $k=J_0^2F(1-\beta_0)B(1-\beta_0, 1+\beta_0)/2$. In the long-time limit, the second term inside the square bracket in Eq. \eqref{eq:int2c} can be neglected and the first term has an asymptotic value of $\pi/2$ for $\beta_0\rightarrow1$. So the long-time behaviour of $\braket{\hat{S}_x(t)}$ is a stretched exponential $\exp[-(\gamma t)^\beta]$ with $\beta=\beta_0$ and $\gamma=J_0[F\cos(\beta_0 \pi/2)\Gamma(1-\beta_0)]^{1/\beta_0}$. As a specific example, we show plots of Eq. \eqref{eq:homo_exp2} with $I(J_0t;\beta_0)$ from Eqs. \eqref{eq:int2a}, \eqref{eq:int2b}, and \eqref{eq:int2c} in Fig. \hyperref[fig:homo]{2(b)} for $\beta_0=0.5$. Other than the two limits discussed before, a damped oscillating decay is observed in between, which to a large extent can be captured by the neglected second term inside the square bracket in Eq. \eqref{eq:int2c}. This oscillating decay signatures the breakdown of scale invariance with the soft-core potential.

\subsection{Gaussian samples: a numerical study} \label{sec:inhomo}

\begin{figure}[t]
	\includegraphics[width=0.5\textwidth]{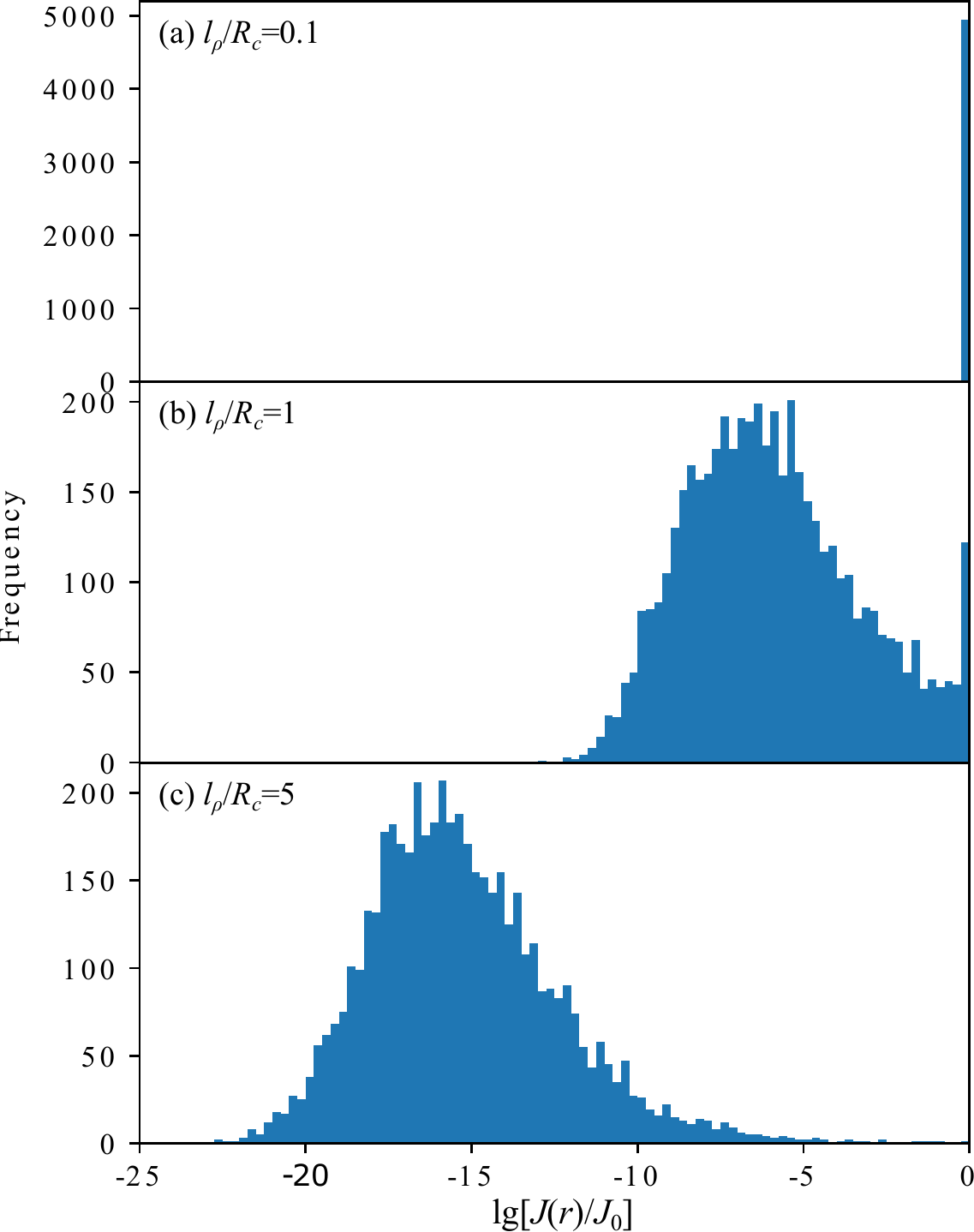}
	\centering
	\caption{Distribution of the pair interaction strength $J(r)$ for 100 spins randomly distributed in a three-dimension Gaussian profile. (a), (b), (c) show the distribution of $\lg[J(r)/J_0]$ for a mean interparticle distance of $l_\rho/R_c=0.1$, 1, 5, respectively.}
	\label{fig:interaction_randomness}
\end{figure}

\begin{figure}[t]
	\includegraphics[width=0.5\textwidth]{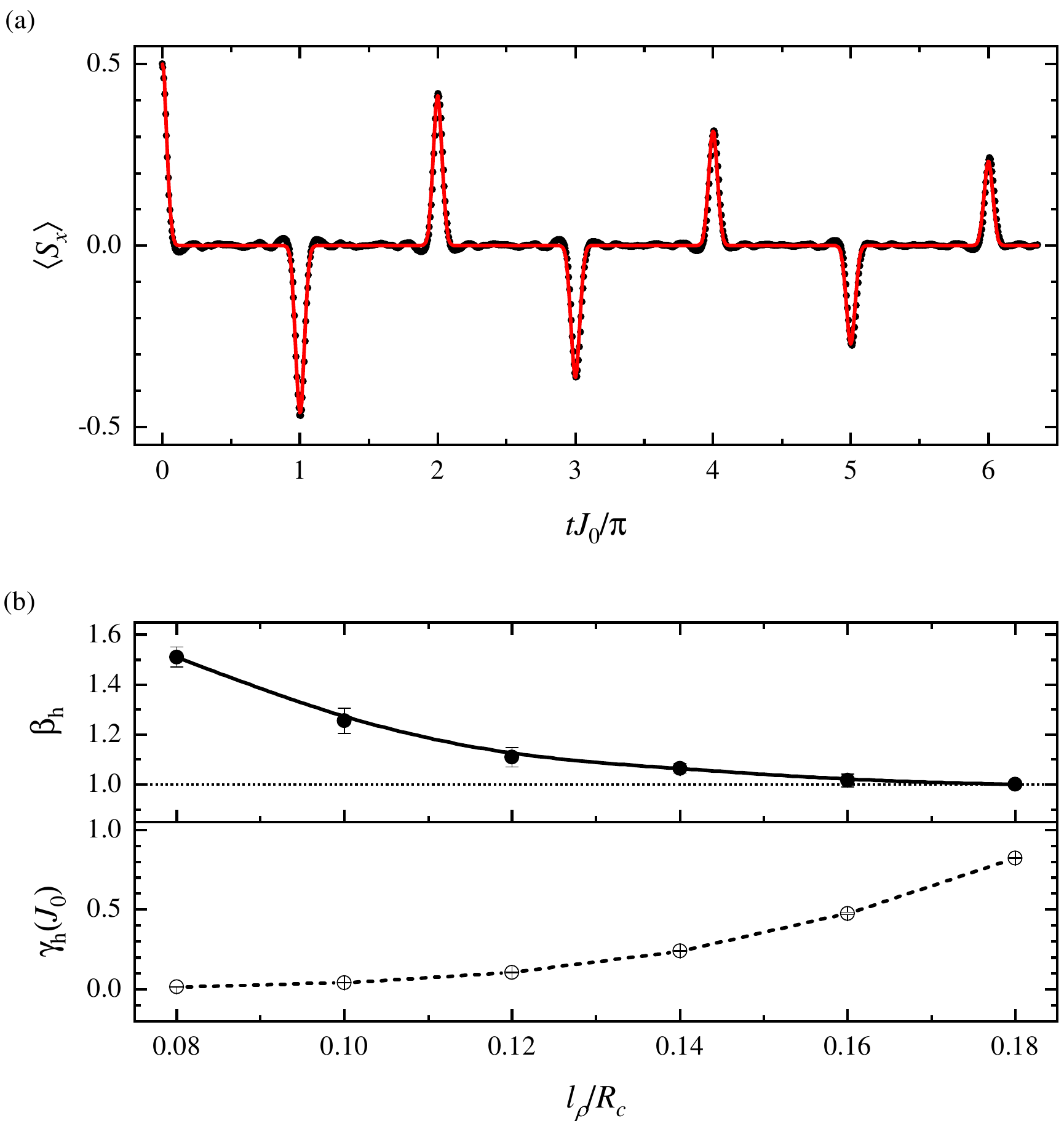}
	\centering
	\caption{Magnetization dynamics in a high-density Gaussian sample. (a) The time evolution of mean magnetization $\braket{\hat{S}_x}$ (solid points) for $l_\rho/R_c=0.1$. After a initial fast decay, quantum revivals are observed with a damped amplitude. The red curve is a fit to the function of $\cos^{N-1}(J_0t)\exp[-(\gamma_ht)^{\beta_h}]$ (see text for details), with which the stretched exponent $\beta_h$ and decay rate $\gamma_h$ are extracted. (c) The fitted $\beta_h$ and $\gamma_h$ as a function of $l_\rho/R_c$. The lines are guides to eyes. See text for more discussions.}
	\label{fig:highdensity}
\end{figure}

To extend the above analytic result for the homogeneous case, we numerically investigate the magnetization relaxation for an inhomogeneously distributed spin sample (Gaussian distributed) in this section, where the degree of disorder can be tuned. We focus on dynamics of the magnetization $\braket{\hat{S}_x(t)}$ under the setting specified in Sec. \ref{sec:hamiltonian}, however, with spin positions $\boldsymbol{r}=(x, y, z)$ randomly distributed in a three-dimension Gaussian distribution ($d=3$)
\begin{equation} \label{eq:gaussian}
	G(\boldsymbol{r}) = \frac{1}{(2\pi)^{3/2}w_x w_y w_z}\exp(-\frac{x^2}{2w_x^2}-\frac{y^2}{2w_y^2}-\frac{z^2}{2w_z^2})\, ,
\end{equation}
where $w_\eta$ is the Gaussian waist in $\eta$ direction ($\eta\in\{x,y,z\}$). This distribution of spins could be realized with ultracold atoms trapped in harmonic traps \cite{Ketterle1999}. The mean particle density is $\rho=N/(8\pi^{3/2}w_x w_y w_z)$ with the total particle number $N$ and for simplicity we assume $w_x=w_y=w_z\equiv w$, giving rise to a mean inter-spin distance of $l_\rho\equiv\rho^{-1/3}=2\sqrt{\pi}w/N^{1/3}$ and its corresponding interaction strength $J_\rho=J(l_\rho)=J_0/[1+(\pi^{d/2}/F\Gamma(d/2+1))^{1/\beta_0}]$ in Eq. \eqref{eq:Vr} with $F, \beta_0$ defined in Sec. \ref{sec:homo}. For following numeric calculation, we fix the total spin number $N=100$ and $\alpha=6$ ($\beta_0=0.5$).

For the soft-core potential in Eq. \eqref{eq:Vr}, $R_c$ separates the interaction-strength randomness into two different regimes according to the ratio $l_\rho/R_c$ for the above Gaussian sample. We show in Fig. \ref{fig:interaction_randomness} the distribution of pair interaction strengths with 100 spins randomly distributed according to Eq. \eqref{eq:gaussian} for three different values of $l_\rho/R_c$: 0.1, 1, and 5. When $l_\rho$ is much smaller than $R_c$ [$l_\rho/R_c=0.1$ in Fig. \hyperref[fig:interaction_randomness]{3(a)}], $J(r)$ is almost the constant $J_0$ for all pairs, hence randomness is minimized. Otherwise when $l_\rho/R_c\gtrsim1$, the distribution of $J(r)$ spans over several orders of magnitude, as seen in Figs. \hyperref[fig:interaction_randomness]{3(b, c)}. Thus effects arising from disorder are expected to be important in this regime. 

\subsubsection{High-density regime}

In Fig. \ref{fig:highdensity} we present the numerical results from Eq. \eqref{eq:EmchRadin} for $l_\rho/R_c\in(0.1, 0.2)$ ($\rho\sim10^{14}-10^{15}$~cm$^{-3}$ for $R_c=1~\mu$m), coined \emph{high-density regime}. In this regime, the system behaves like a all-to-all interacting one with a single interaction strength $J_0$ \cite{Schachenmayer2015}, recovering a coherent many-body dynamics, as seen in Fig. \hyperref[fig:highdensity]{4(a)} for the magnetization dynamics with $l_\rho/R_c=0.1$. A fast initial decay of magnetization due to the buildup of correlations \cite{Bohnet2016, Zeiher2017} and periodic quantum revivals with decaying amplitudes are observed.

In Eq. \eqref{eq:EmchRadin}, if all the interaction strengths take a common value of $J_0$, the resulting dynamics of $\braket{\hat{S}_x(t)}$ has an analytic form, $\braket{\hat{S}_x(t)}=\cos^{N-1}(J_0t)$, giving rise to a coherent many-body quantum-revival dynamics \cite{Schachenmayer2015}. To account for the observed decaying revival dynamics in Fig. \hyperref[fig:highdensity]{4(a)}, we phenomenologically fit the numerical data to a form of $\cos^{N-1}(J_0t)\exp[-(\gamma_ht)^{\beta_h}]$, which is a stretched-exponential decay and shown as the red curve in the figure. Note that we have tried fits with a pure exponential decay, which can not fully capture the observed dynamics. From the fit, we obtain the stretched exponents $\beta_h$ and decay rates $\gamma_h$ for various values of $l_\rho/R_c$, which are plotted in Fig. \hyperref[fig:highdensity]{4(b)}. We observe a monotonic approach to the normal exponential decay ($\beta_h=1$) from $\beta_h>1$ with an increasing disorder in the system (see Fig. \ref{fig:interaction_randomness}), while the decay rate $\gamma_h$ increases from 0 to the order of $J_0$.

\subsubsection{Low-density regime}

\begin{figure}[t]
	\includegraphics[width=0.5\textwidth]{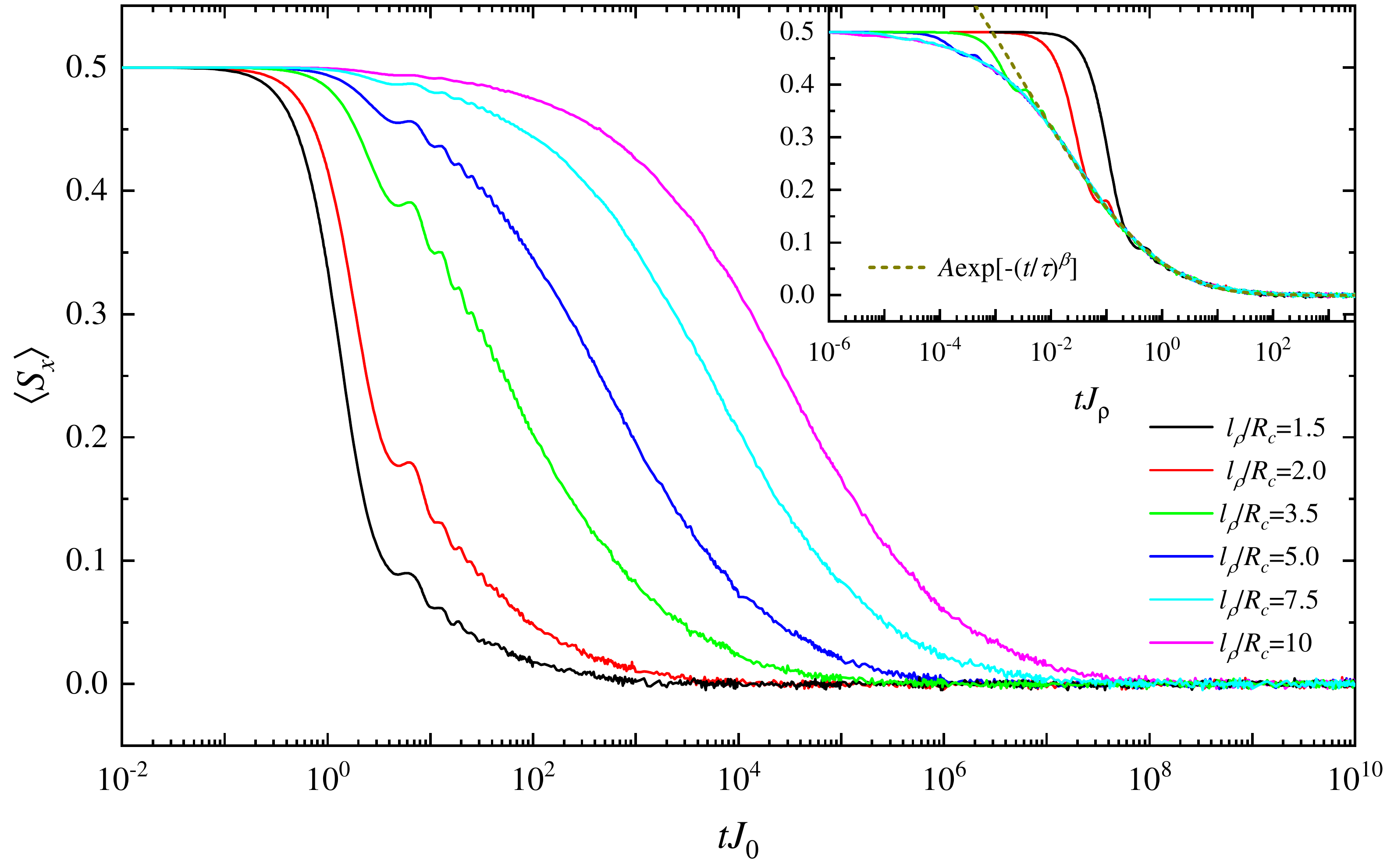}
	\centering
	\caption{Magnetization dynamics in a low-density Gaussian sample. Evolution curves of $\braket{\hat{S}_x}$ for six different values of $l_\rho/R_c$ ranging from 1.5 to 10 are plotted. As $l_\rho$ increases ($\rho$ decreases), the initial fast decay seen in Fig. \hyperref[fig:highdensity]{4(a)} shrinks along the time axis and a slow dynamics similar to that in Fig. \ref{fig:homo} emerges. In the inset, the rescaled evolution curves ($tJ_0\rightarrow tJ_\rho$) are shown, all of which fall onto a common one at the long-time part. We fit this common part to a stretched-exponential function (dashed dark yellow curve) and obtain an exponent of $\beta=0.1817(9)$ and a decay rate of $\gamma=343(15)J_\rho$.}
	\label{fig:lowdensity}
\end{figure}

In the other regime with $l_\rho/R_c>1$ ($\rho<10^{12}$~cm$^{-3}$ for $R_c=1~\mu$m), the pair interaction strengths distribute over a range covering 5 or 6 orders of magnitude (see Fig. \ref{fig:interaction_randomness}). The results for dynamics of $\braket{\hat{S}_x}$ at six different values of $l_\rho/R_c$ are shown in Fig. \ref{fig:lowdensity}. As the cloud size $w$ ($l_\rho$) increases, the initial collapse phase [see Fig. \hyperref[fig:highdensity]{4(a)}] shrinks and a slow decay with an oscillating feature merges at long time, which is similar as that in Fig. \hyperref[fig:homo]{2(b)}. 

We rescale the time $t$ for each curve in Fig. \ref{fig:lowdensity} by the characteristic interaction strength $J_\rho$ (as introduced in the beginning of this section), which is presented in the inset. The curves for different $l_\rho/R_c$ fall onto a common one at the long-time part including the oscillation, which occupies a larger region in the dynamics for larger $l_\rho/R_c$ and demonstrates a universal behavior. We fit this common long-time part of the $\braket{\hat{S}_x}$ dynamics to the stretched-exponential function $A\exp[-(\gamma t)^\beta]$ with the fitting parameter $A, \gamma, \beta$ [dashed curves in the inset of Fig. \ref{fig:lowdensity}. The fitted exponent $\beta$ is 0.1817(9) and the decay rate $\gamma$ is $343(15)J_\rho\approx19.5(9)J_0F^{1/\beta_0}$ ($F\ll1$). Both the stretched exponent and decay rate are different from the values obtained analytically for a homogeneous sample with $\beta_0=0.5$ in Sec. \ref{sec:homo}, where $\beta=\beta_0=0.5, \gamma\approx1.57J_0F^{1/\beta_0}$.

\section{Conclusion} \label{sec:conclusion}
In conclusion, we have considered magnetization relaxation of homogeneous and inhomogeneous samples of Ising spins with a soft-core pairwise potential. We have derived an analytic formula describing the whole dynamics in the homogeneous case, with three distinct relaxation regions in the time axis. The short-time dynamics follows $\exp(-kt^2)$ and stretched-exponential laws are found at long-time dynamics. As conjectured by Klafter and Shlesinger, this law arises from a scale-invariant distribution of relaxation times, which is only approximately fulled in the long-time limit since the soft-core potential in general is not scale-invariant. The breakdown of scale invariance is indicated by an oscillating feature in the relaxation between the short- and long-time limit. 

Similar behaviours emerge for large Gaussian samples compared to the soft-core radius, where strong disorder presents in the system. However, for small Gaussian samples a coherent many-body dynamics is found since all spins interact with each other with an almost constant interaction strength. A smooth change from the coherent regime to the strongly disordered regime can be realized via tuning the Gaussian size of the sample. Our results in both homogeneous and inhomogeneous situations may stimulate experimental interests in the cold-atom community and may also be generalized to other types of interaction potentials.

\section*{Acknowledgements}
We are grateful to Xiaopeng Li and the Rydberg team of Weidem\"uller's group in Heidelberg for careful reading of our manuscript. We are supported by the Anhui Initiative in Quantum Information Technologies. Y.H.J. also acknowledges support from the National Natural Science Foundation under Grant No. 11827806. M.W. is supported by the Deutsche Forschungsgemeinschaft (DFG, German Research Foundation) under Germany’s Excellence Strategy EXC2181/1-390900948 (the Heidelberg STRUCTURES
Excellence Cluster), within the Collaborative Research Center SFB1225 (ISOQUANT) and the DFG Priority Program 1929 "GiRyd" (DFG WE2661/12-1).

\vspace{2em}
\section*{Appendix}

\subsection*{Analytic derivation in a homogeneous sample} \label{app:ad_homo}

To derive Eq. \eqref{eq:int2a} from $I(J_0 t; \beta_0)=\int_{y_0}^{J_0t} (J_0t/y-1)^{\beta_0}\sin y dy$ for $\beta_0<1$, we first introduce a new variable $x=1/y$ in the later integral, resulting in
\begin{widetext}
\begin{equation} \label{eq:homo_exp3}
	\begin{aligned}
		I(J_0 t; \beta_0) & =\int_{1/(J_0t)}^{1/y_0} (J_0tx-1)^{\beta_0}x^{-2}\sin(x^{-1}) dx \\
		\xRightarrow{y_0\rightarrow0} &  \frac{(J_0t)^{\beta_0}}{2i}\int_{1/(J_0t)}^{\infty} (x-\frac{1}{J_0t})^{\beta_0}x^{-2}(e^{ix^{-1}}-e^{-ix^{-1}}) dx \\
		& = \frac{(J_0t)^{\beta_0}}{2i}B(1-\beta_0, 1+\beta_0)(J_0t)^{1-\beta_0}[ _1F_1(1-\beta_0,2,iJ_0t)-\text{ } _1F_1(1-\beta_0,2,-iJ_0t)] \\
		& =J_0tB(1-\beta_0, 1+\beta_0)\Im[_1F_1(1-\beta_0,2,iJ_0t)]
	\end{aligned}
	\, .
\end{equation}
\end{widetext}
Here we have used a integral formula listed in Ref. \cite{Gradshtein2015}, which reads
\begin{widetext}
\begin{equation} \label{eq:integ}
	\begin{aligned}
		\int_m^\infty x^{v-1}(x-m)^{\mu-1}e^{b/x}dx=B(1-\mu-v,\mu)m^{\mu+v-1} _1F_1(1-\mu-v,1-v,b/m)
	\end{aligned}
	\, ,
\end{equation}
\end{widetext}
and is valid for $m>0, 0<\Re(\mu)<\Re(1-v)$. $\Re(z)$ represents the real part of $z$. The asymptotic behavior of the Kummer confluent hypergeometric function $_1F_1(a,b,z)$ at large $|z|$ is $_1F_1(a,b,z)\sim\Gamma(b)[e^z z^{a-b}/\Gamma(a)+(-z)^{-a}/\Gamma(b-a)]$, which gives rise to Eq. \eqref{eq:int2c}.

\bibliography{mylibrary}

\end{document}